# Variance-based Distributed Clustering

Lamine M. Aouad, Nhien-An Le-Khac, and Tahar Kechadi
University College Dublin, School of Computer Science & Informatics
Belfield, Dublin 4 – Ireland
{lamine.aouad,an.le-khac,tahar.kechadi}@ucd.ie

**Abstract**

*Nowadays, huge amounts of data are naturally collected in distributed sites due to different facts and moving these data through the network for extracting useful knowledge is almost unfeasible for either technical reasons or policies. Furthermore, classical parallel algorithms cannot be applied, specially in loosely coupled environments. This requires to develop scalable distributed algorithms able to return the global knowledge by aggregating local results in an effective way. In this paper we propose a distributed algorithm based on independent local clustering processes and a global merging based on minimum variance increases and requires a limited communication overhead. We also introduce the notion of distributed sub-clusters perturbation to improve the global generated distribution. We show that this algorithm improves the quality of clustering compared to classical local centralized ones and is able to find real global data nature or distribution.*

## 1 Introduction

Clustering is one of the fundamental technique in data mining. It groups data objects based on information found in the data that describes the objects and their relationships. The goal is to maximize similarity within a group and the difference between groups in order to identify interesting distributions in the underlying data. This is a difficult task in unsupervised knowledge discovery and there is a large amount of literature in the subject ranging from models, algorithms, validity and quality studies... However, there is still several issues in the clustering process including how to find the optimal number of clusters, how to assess the validity of a given clustering, how to allow different and naturel shapes and sizes rather than forcing them into normed balls of the distance function, how to prevent the algorithms initialization and the order in which the features vectors are read in from affecting the clustering output, and how to find which clustering structure in a given dataset, i.e why choosing a given algorithm instead of another. Most of these issues comes from the fact that there is no general definition of what is a cluster. In fact, algorithms have been developed to find several kinds of clusters; spherical, linear, dense, drawnout, etc.

Furthermore, in a distributed way, clustering algorithms will deal with the problem of distributed data, computing nodes and domains, plural ownership and users, and scalability. On the other hand, moving the entire data to a single location for processing could be impossible due to different reasons related to policies or technical choices. Also, the communication efficiency of an algorithm is often more important than the accuracy of its results. In fact, communication issues are a key factor in the implementation of distributed algorithms. It is obvious that a suitable algorithm for high speed network can be of little use in WAN-based one. We considere that an efficient distributed algorithm need to exchange a few data and avoid synchronization as much as possible.

In this paper, we propose a distributed algorithm that forms global clusters based on sub-clusters merging variance constraint. This improves the overall clustering quality and can find the number of clusters automatically. However, a proper maximum increasing value has to be selected. This can be deducted from the problem domain or found out using various methods. We will briefly present the way presented in [23] in the experiments section.

This study is part of a distributed data mining project called ADMIRE [24]. The rest of the paper is organized as follows, the next section surveys some previous parallelisation and distribution efforts in the clustering area. Then, section 3 presents our distributed algorithm. Section 4 shows some experimental results and evaluations and highlights directions for future work. Finally, section 5 gives the conclusion.

## 2 Related Work

This section survey some works in parallel and distributed clustering. We also discuss the latest projects and propositions especially regarding grid-based approaches.

First, clustering algorithms can mainly be separated into two general categories, namely partitioning and hierarchical. Different elaborated taxonomies of clustering algorithms are given in the literature [5]. An overview of this general classification scheme, with some important algorithms, is given in Fig. **??**. Details about these algorithms is out of the purpose of this paper, we refer the interested reader to [5] and [8] where good overviews and further references can be found.

Many parallel clustering algorithms have been considered [13][14][16][15]... In [13] and [14], a message-passing versions of the widely used k-means algorithm were proposed. In [16] and [15], the authors deal with the parallelisation of the DBSCAN density based clustering algorithm. Most of these approaches need either (or both) a global view of the dataset or multiple synchronization constraints between processes.

The distributed approach is different, even many of the proposed distributed algorithms are based on algorithms which were developed for parallel systems. Actually, most of them typically act by producing local models followed by the generation of a global model through the aggregation of the local results in different ways. This assumes a kind of *peer-to-peer way* in the sens that the processes participating to the computation act independently and have the same computation level, i.e can be server or client. According to this, the global clustering is established based on only local models, without a global view. All these algorithms are then based on the global reduction of so-called sufficient statistics, probably followed by a broadcast of the result. Most notable examples are the works presented in [3][4][9][17][18], mostly related to the k-means algorithm or variants and the DBSCAN density based algorithm.

On the other hand, grid and peer-to-peer systems have emerged as an important area in distributed and parallel computing[1]. In the data mining domain, where massive datasets are collected and need to be stored and performed, the grid can be seen as a new computational and large-scale support, and even as a high performance support in some cases. Some grid or peer-to-peer based projects and frameworks already exist or are being proposed in this area; Knowledge Grid [22], Grid Miner [25], Discovery Net [26], ADMIRE [24]... Beyond the architecture design of these systems, the data analysis, integration or placement approaches, the underlying middleware and tools, etc. the grid-based approach needs efficient and well-adapted algorithms. This is the motivation of this work.

## 3 Algorithm description

This section describes our distributed algoritm and gives some formal definitions. The key idea of this algorithm is to choose a relatively high number of clusters locally (which will be called sub-clusters in the rest of the paper) and to merge them at the global level according to an increasing variance criterion which require a very limited communication overhead. All local clustering are independent from each other and the global merging can be done independently, from and at any initial local process.

### 3.1 Algorithm foundations

At the local level, the clustering can be done by different clustering algorithms; k-means, k-harmonicmeans, k-medoids, or variants, or using the statistical interpretation with the expectation-maximization algorithm which finds clusters by determining a mixture of Gaussians distributions. The merging process of local sub-clusters at the global level exploits locality in the feature space, i.e., the most promising candidates to form a global cluster are sub-clusters that are the closest in the feature space.

---
[1]Again, the designation 'parallel' is used here to highlight the fact that the computing tasks are interdependent, which is not necessarily the case in distributed computing.

Each participating process can execute the merging actions and substract the global cluster formation, i.e., which sub-clusters are susceptible to form together a global one.

Before describing the algorithm itself, we first give developments on some used notions. A global cluster border represents local sub-clusters at its border. These are susceptible to be isolated and added to another global cluster in order to contribute to an improvement of the merging constraint. These sub-clustres are referred as perturbation candidates. Actually, the initial merging order may affect the clustering output, this action is intended for minimizing the input order impact. The global clusters are then updated. The border is collected by computing the common Euclidean distance measure. The k farthest sub-clusters are then our candidates, where k is a predefined number. This value depends on the chosen local number of clusters. Furthermore, multi-attributed sub-clusters are naturally concerned by perturbation.

The algorithm starts with $\sum_{i \in s} k_i$ clusters, where s is the number of sites involved and $k_i$, $i \in s$ the local chosen number of clusters on s. Then, each process has the possibility to generate the global merging. An important thing here is that the merging is logical, i.e each local process can generate correspondances between local sub-clusters, without necessarily reconstructing the overall clustering output. That is because the only bookkeeping needed from the other sites are centers, counts and variances. The perturbation process acts finally if the merging action is no longer applied. k candidates are collected for each global cluster from its border, which is proportional to the overall size composition, as quoted before. Then, the process move these candidates by trying the closest ones and with respect to the gain in the variance criterion when moving them from the neighboring global clusters. In the next section we will formally define the problem, notions and criterions.

## 3.2 Formal definitions

This section formalize the clustering problem and the notions described in the previous section. Let $X = \{x_1, x_2, ..., x_N\}$ be a data set of N elements in the p-dimensional metric space. The clustering problem is to find a clustering of X in a set of clusters $C = \{C_1, C_2, ..., C_M\}$ such that the clusters are homogeneous. The most used criterion to quantify this homogeneity is the variance criterion, or sum-of-squared-error (SSE) criterion :

$$S = \sum_{i=1}^{M} SSE(C_i)$$

where

$$SSE(C) = \sum_{x \in C} \|x - u(C)\|^2$$

and

$$u(C) = \frac{1}{|C|} \sum_{x \in C} x$$

is the cluster mean.

Traditional constraints used to minimize the given criterion is to fix the number of clusters M to an a priori known number as in the widely used k-means, k-harmonicmeans, k-medoids algorithms or variants (CLARA, CLARANS)...[5][6][7]. This constraint is a very restrictive one since this number is most likely not known in most cases. However, many estimating techniques exist in the literature, as the gap statistic which compare the change in within cluster dispertion to that expected under an appropriate reference null distribution [27] or the index due to Calinski & Harabasz [28], etc... These methods need a global view of the data and no distributed version exists. Oppositely, the imposed constraint here states that the increasing variance of the merging, or union, of two clusters is below a limit $\sigma_{i,j}^{max}$ defined as twice the highest individual variance from sub-clusters $C_i$ and $C_j$ [23].

The border $B_i$ of the cluster $C_i$ is the set of the k fartest sub-clusters from the new global cluster center (considering at this time $C_i$ as a global cluster). Let $SC_i = \{scc_1, scc_2, ..., scc_{n_i}\}$ be the set of the $n_i$ sub-clusters centers merged into $C_i$. $B_i$ is defined as :

$$B_i(k) = F(c_i^{new}, k, C_i, SC_i)$$

where

$$F(c_i^{new}, k, C_i, SC_i) = \begin{cases} fsc(c_i^{new}, k, C_i, SC_i) \cup F(c_i^{new}, k-1, C_i, SC_i - fsc(c_i^{new}, k, C_i, SC_i)), & k > 0 \\ \emptyset, & k = 0 \end{cases}$$

$fsc(c_i^{new}, k, C_i, SC_i)$ are the k farthest sub-clusters' centers from $c_i^{new}$ :

$$fsc(c_{new}, k, C_i, SC_i) = \arg\min_{x \in SC_i} Euclidean(x, c_i^{new})$$

These sets are then performed once the merging is no longer applied and as quoted before, the multi-attributed sub-clusters should belong to it.

## Similarity metrics

Additionally, the merging process could needs to calculate a 'distance' between local sub-clusters. Then, everyone could be described by additional objects and metrics, as the covariance matrices for example. A general definition of a distance measure can be $d(x_i, x_j) = (c_j - c_i)^T A (c_j - c_i)$, where the inclusion of A results in weighting according to statistical propreties of the features. Other possible distance/similarity measures include; Euclidean, Manhattan, Canberra, Cosine, Squared chord, Squared Chi-squared, Chebychev.. The general form of some of these distances is $d_{i,j} = [\sum_K |x_{ki} - x_{kj}|^N]^{\frac{1}{N}}$, and depending on N, the enclosed region takes different shapes. That is to say that the merging process could take into acount one or different proximity (i.e similarity or dissimilarity) function to improve the quality of the resulting clustering. This will be considered in future versions. However, the key issue is the selection of the 'right' function, especially, *which kind of measures for which kind of data ?*

## summarized algorithm

According to the previous definitions and formalism, here is the algorithm :

INPUT : s datasets $X_i$, and $k_i$, the number of sub-clusters in each node i
OUTPUT : $k_{global}$ global clusters, i.e the global sub-clusters distribution

### Variance based distributed clustering algorithm

1. Perform local clustering in each local data set, $k_i$ is relatively large (and can be different in each site). Each local clustering gives as output sub-clusters identified by a unique identifier $Cluster_{\{process,number\}}$, and their sizes, centers and variances.

   - For i = 0..s
     cluster($X_i$,$k_i$)

2. At the end of local processes, local statistics are sent to a chosen merging process j.

   - For i = 0..s
     send(j,$sizes_i$,$centers_i$,$variances_i$), i /= j

3. Merging sub-clusters in two phases :

   - While ($Var(C_i, C_j) < \sigma_{i,j}^{max}$)
     merge($C_i$,$C_j$)
   - *perturbation* :
     - Find $B_i(k)$, $i \in k_{global}$ and k is a user defined parameter, and adding multi-attributed sub-clusters if not already in,
     - for each $x \in B_i(k)$, find the closet global cluster j and compute the new variance; $Var_{total}(C_i - C_x, C_j + C_x)$.

At the phase 1 of the merging process, the new global statistics are :

$$N_{new} = N_i + N_j$$

$$c_{new} = \frac{N_i}{N_{new}} c_i + \frac{N_j}{N_{new}} c_j$$

$$v_{new} = v_i + v_j + inc(i, j), \forall C_i, C_j, i /= j$$

when

$$inc(i, j) = \frac{count_i \times count_j}{count_i + count_j} \times Euclidean(C_i, C_j)$$

represents the insreasing in the variance while merging $C_i$ and $C_j$.

As in all clustering algorithms, the potentiel large variability in clusters shapes and densities is an issue. However, as we will show in the next section, the algorithm is efficient to detect well separated clusters and distribution with their effective distribution number. Otherwise, a clear definition of a cluster does not exist anymore. This is also an efficient way to improve the output for a k-mean clustering for example, without an a priori knowledge about the data or an estimation process for the number of clusters.

## Analysis of performance

The computational complexity of this distributed algorithm depends on the algorithm used locally, the communication time (a gather operation) and the merging computing time :

$$T = T_{comp} + T_{comm} + T_{merge}$$

If the local clustering is a k-means, $T_{comp} = O(N_i k_i d)$, where d is the dimension of the dataset. The communication time is the reduction of $3d \sum_{i_s} k_i$ elements. Actually, one process (let us say where the global information is needed) gathers these information in order to perform the merging. If $t^i_{comm}$ is the communication cost for one element from the site i then $T_{comm} = 3d \sum_{i_s} t^i_{comm} k_i$. Since $k_i$ is usually much less large than $N_i$, the generated communication overhead is very limited.

The merging process is executed a number of times, say u. This integer is the number of iterations until the condition $Var(C_i, C_j) < \sigma_j^{max}$ is no longer applied. This cost is then $u \times t_{newStatistcs}$, which corresponds to O(d). This is followed by a perturbation process, which the cost is of order $O(k k_{global} k_i)$, since this process computes for each of the k chosen sub-cluster at the border of i, $k_i$ distances for each of the $k_{global}$ global clusters. The total cost is then :

$$T = O(N_i k_i d) + O(d) + O(k k_{global} k_i) +$$
$$T_{comm}, \quad T_{comm} << O(N_i k_i d)$$

## 4 Experiments

In this section we show the efficiency of the proposed algorithm with some artificial and real data sets. First, we give a description of the data and the experimental environment and tools, and briefly explain how to set up the constraint parameter, i.e the maximum merging variance as twice the highest individual sub-cluster variance.

### 4.1 Data description

The first dataset is a generated random Gaussian distributions with 1150 two-dimensional samples. Fig. 1 displays this dataset. The data was randomly distributed in three sets as shown in Fig. 2. The second one is the well-known Iris dataset. It consists in three classes of irises (Iris setosa, Iris versicolor and Iris virginica) each characterized by 8 attributes and there is 150 instances. The set was randomly distributed as shown in Fig. 4 (presented by the attributes "sepal area" and "petal area"). This figure shows also the initial local clustering using k-harmonicmeans with $k = 5$.

### 4.2 Evaluations and discussion

The merging output of the first dataset is shown in Fig. 3. This result find the 'right' number of cluster and their distribution independently of the local used clustering algorithm and the number of clusters. The expectation-maximisation and the k-harmonicmeans algorithms give the same output. The resulting global clustering for the Iris dataset (and a global k-harmonicmeans clustering using the entire dataset) are given is Fig. 5. The algorithm manages to find the class distribution of the Iris dataset, leading to 3 classes based on 5 local sub-clusters. However, because the k-harmonicmeans does not impose a variance constraint it could find a lower sum-of-squared-error which is the case here. These two examples show the independence from the nature and size of the initial clustering. Actually, if there is a 'real' structure in the dataset then true clusters are found and joined together.

In contrast to many other distributed algorithm, the presented one uses a simple global constraint, a very limited communication overhead, and does not need to know the data structure a priori. This algorithm is effective in finding proper clustering, however, future versions will take into account some other facts as considering the perturbation process during the merging operations and inside sub-clusters, or of whether or not multi-attributed clusters are present to consider a different approach at this level. Also, varying the constraint criterion could be considered and the addition of other similarity functions as quoted before.

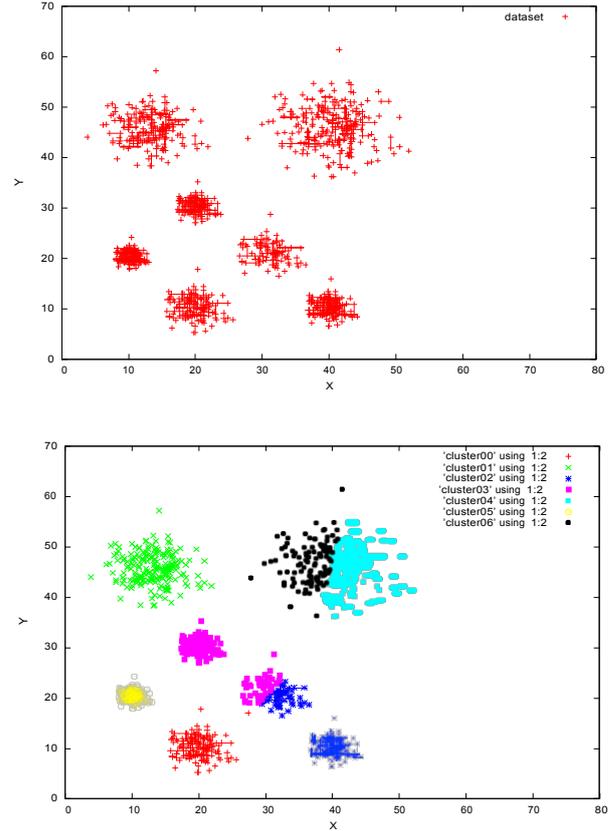

**Figure 1.** The dataset and a global k-means, $k = 7$.

## 5 Conclusion

In this paper, we evoked the need of efficient distributed and grid-based clustering algorithms. Actually, a huge effort has been made in sequential clustering but there is only few algorithms which tackle this problem. We proposed a distributed algorithm based on a variance constraint. It clusters the data locally and independently from each other and only limited statistics about the local clustering are transmitted to the aggregation process which carries out the global clustering, defined as labeling between sub-clusters, by means of a merging and a perturbation processes. The global model can then be broadcasted to all processes if needed, which use it to label their sub-clusters.

The algorithm gives good performances at identifying well separated clusters and the real structure of the dataset. In fact, when data are not well separated, the notion of cluster is very confused and does not even exist in the literature. The number of clusters is also automatically found, this can resolves the problem of estimating the number of clusters a priori. Furthermore, in addition to classical constraints in distributed clustering related to centralising the data due to technical or security reasons or local policies, this algorithm can also tackle large and high dimensional datasets that cannot fit in memory since most of the clustering algorithms in literature require the whole data in the main memory. Nevertheless, open issues could be considered as in the merging process or the choice of the possible better local models, in addition to those described in the previous section.

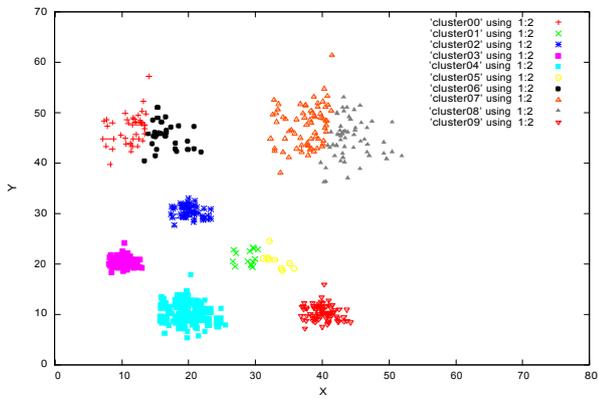

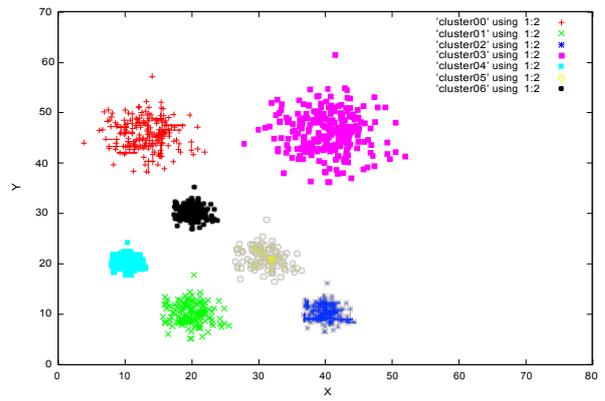

**Figure 3.** Generated clustering.

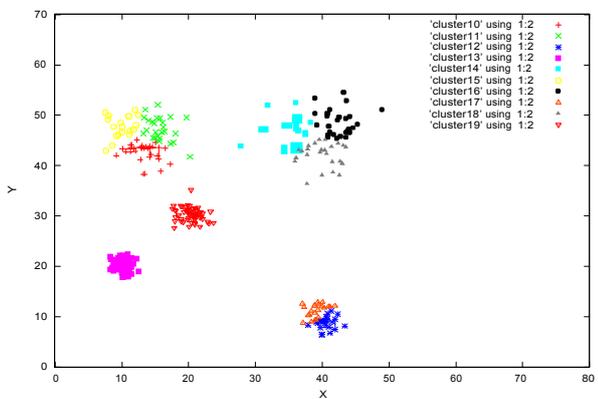

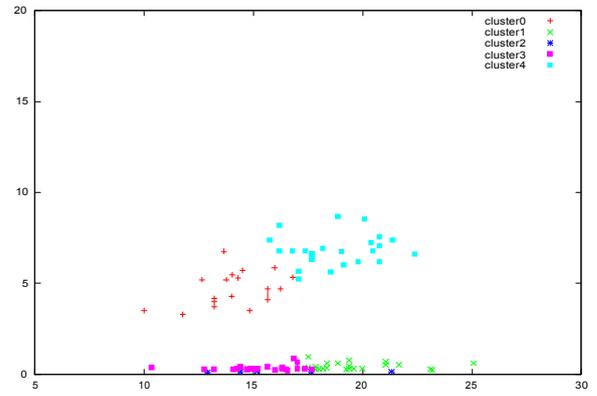

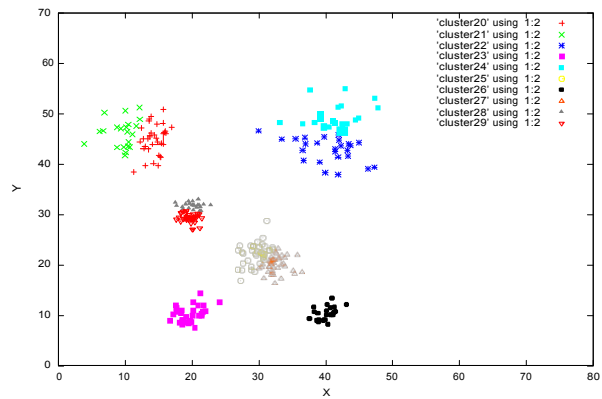

**Figure 2.** Local k-means, $k_i = 10$, $i = 0..2$.

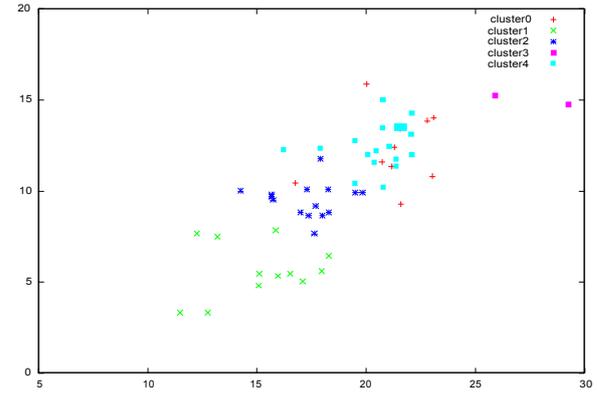

**Figure 4.** Iris sub-sets and local clustering using k-harmonicmeans, $k_i = 5$, $i = 0, 1$.

## Acknowledgments

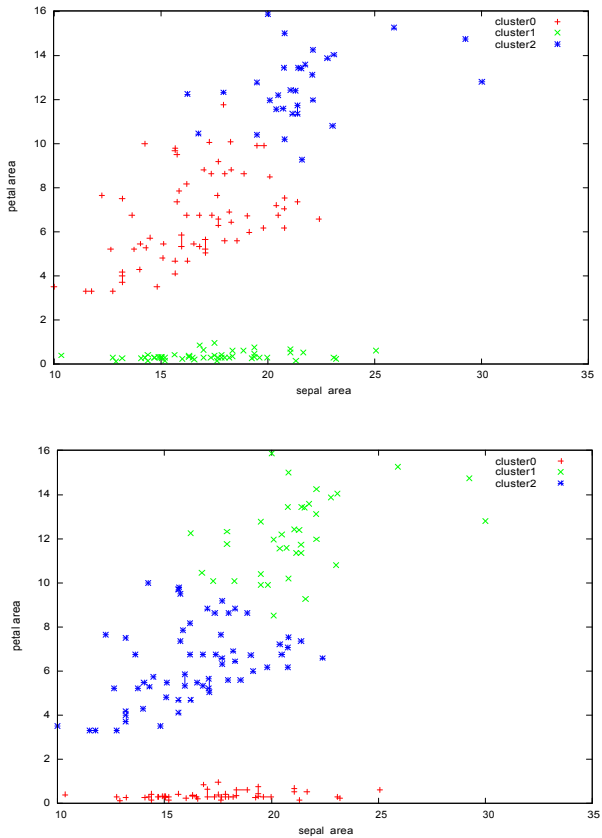

**Figure 5.** The distributed algorithm output and a global clustering using k-harmonicmeans.